\input{aipcheck}

\documentclass[
    ,final            % use final for the camera ready runs
  ]
  {aipproc}

\layoutstyle{6x9}

\usepackage{natbib}

\begin{document}

\title{Intermittent accreting millisecond pulsars:\\ light houses with broken lamps?}



\author{D. Altamirano}{}

\author{P. Casella}{address={University of Amsterdam}}

\begin{abstract} Intermittent accreting millisecond X-ray pulsars are an exciting new
  type of sources. Their pulsations appear and disappear either on
  timescales of hundreds of seconds or on timescales of days. The
  study of these sources add new observational constraints to present
  models that explain the presence or not of pulsations in neutron
  star LMXBs. In this paper we present preliminary results on spectral
  and aperiodic variability studies of all intermittent AMSPs, with a
  particular focus on the comparison between pulsating and non
  pulsating periods.

\end{abstract}

\maketitle

%%%%%%%%%%%%%%%%%%%%%%%%%%%%%%%%%%%%%%%%%%%%
%% MAINMATTER
%%%%%%%%%%%%%%%%%%%%%%%%%%%%%%%%%%%%%%%%%%%%

\section{Introduction}

Already in the early 1980s it was predicted \citep{Alpar82,Backer82}
that the progenitors of millisecond radio pulsars should be
accretion-powered millisecond X-ray pulsars (AMSPs) in neutron-star
low-mass X-ray binaries (LMXBs). However, it was not until 1998 that
the first AMSP was discovered \citep{Wijnands98}. Since then, a total
of seven AMSPs have been found but still for most neutron-star LMXBs
no coherent millisecond X-ray pulsations are detected.  Recently
three sources were found which showed characteristics connecting these
two types of behavior.

HETE~J1900.1-2455 \citep{Kaaret06} is a transient source which has
remained active for $\sim3$~years\footnote{At the time of submitting
  this contribution, the source is still active -- see also Duncan
  Galloway's contribution.} but (intermittent) pulsations were
detected during only the first $\sim70$ days of activity
\citep[see][]{Galloway07a}. On three occasions during this period, it
was observed an abrupt increase in the pulse amplitude, approximately
coincident with the time of a thermonuclear burst, followed by a
steady decrease on a timescale of ~10 days. Although there seems to be
a relation between between occurrence of X-ray burst and onset of
pulsations, only 1 X-ray burst was observed with PCA, while the other
7 were observed with Swift or HETE-II. Therefore, the delay between the burst and the onset of pulsations is not always known. No burst oscillations have been detected yet.

Aql~X-1 is a $\sim550$~Hz burst oscillation transient source from
which pulsations were detected \citep{Casella08} only for
$\sim150$~sec out of the $\sim1.5$~Msec the source has so far been
observed with RXTE. The pulsations were discovered at the end of a
one-orbit $\sim1500$-seconds observation. Given the data structure, we
cannot exclude the occurrence of an X-ray burst $\sim1400$ seconds
before the pulsations or immediately after the pulsations disappeared
\citep[see Figure 2 in][]{Casella08}.

SAX~J1748-2021 is a transient source from which intermittent
pulsations were detected during the 2001 and 2005 outbursts
\citep{Altamirano08b}\footnote{We emphasize that this is the third AMSPs
showing more than a single outburst; the other two are
SAX~J1808-.4-3658 and IGR~J00291+5934.}. The pulsations appeared and
disappeared on timescales of hundreds of seconds. A suggestive
relation between the occurrence of type-I X-ray bursts and the
appearance of the pulsations was found, but the relation is not strict
\citep{Altamirano08b,Patruno07}. Similarly to HETE~J1900.1-2455, no
burst oscillations have been detected.

It is unclear if the intermittence of the pulsations in these three
sources are caused by the same mechanism and if they are unique among
the neutron-star LMXBs. In this paper, we present preliminary results
on our effort to further explore on the differences and similarities
between AMSPs, intermittent AMSPs and non-pulsating neutron star
systems.

\section{Source states and pulsations}

\citet{Hasinger89} classified the NS LMXBs based on the correlated
variations of the X-ray spectral and rapid X-ray variability
properties. They distinguished two sub-types of NS LMXBs, the Z
sources and the atoll sources, whose names were inspired by the shapes
of the tracks they trace out in an X-ray color-color diagram on time
scales of hours to days. The Z sources are the most luminous while the
atoll sources cover a much wider range in luminosities
\citep[e.g. ,][and references therein]{Ford00}.
For each type of source, several spectral/timing states are identified
which are thought to arise from qualitatively different inner flow
configurations.  In the case of atoll sources, the main three states
are the extreme island state (EIS), the island state (IS) and the
banana branch (BB). Each state is characterized by a unique
combination of spectral and timing behavior.

\begin{figure}[!hbtp] 
\resizebox{1\columnwidth}{!}{\rotatebox{0}{\includegraphics[clip]{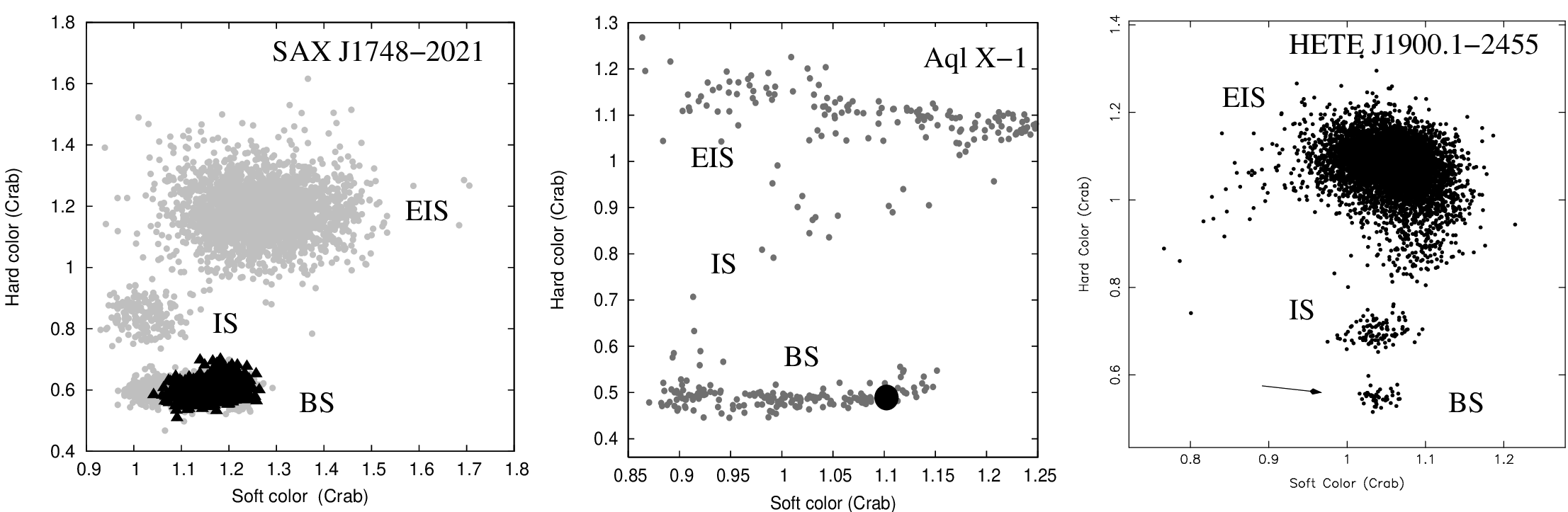}}}
%\resizebox{0.3\columnwidth}{!}{\rotatebox{-90}{\includegraphics[clip]{ccd.ps}}}
%\resizebox{0.3\columnwidth}{!}{\rotatebox{-90}{\includegraphics[clip]{ccdqxl.ps}}}
%\resizebox{0.3\columnwidth}{!}{\rotatebox{0}{\includegraphics[clip]{hete_ccd.eps}}}
%\resizebox{0.4\columnwidth}{!}{\rotatebox{-90}{\includegraphics{hid.ps}}}
%\resizebox{0.4\columnwidth}{!}{\rotatebox{-90}{\includegraphics{hidaql.ps}}}
\caption{Color Color diagrams for SAX J1748--2021 (16s averages), Aql
  X-1 (average per observation) and HETE~J1900.1-2455 (16s
  averages). The black triangles and the circle show the source state
  when pulsations in SAX J1748--2021 and Aql X-1 were present. The
  arrow marks the detection of kHz QPOs in HETE~J1900.1-2455. This
  detection plus the colors confirm that this source was in its soft
  state. }\label{fig:ccd}
\end{figure}

Interestingly, all AMSPs with persistent pulsations have been observed
\textbf{only} during their island or extreme island state
\citep{Wijnands05}. The 3 intermittent AMSPs are the only pulsating
sources that have been observed not only in their island states, but
also in their (soft) banana states. In Figure~\ref{fig:ccd} we show
the color-color diagrams for these three sources. For Aql X-1 and
SAX~J1748--2021 we also mark where the pulsations were found. For
HETE~J1900.1-2455, the pulsations are only present in the EIS/IS.

\section{Power spectra in the Intermittent AMSPs}

It has been shown that the frequencies of all the components (except
for those of the so-called hectoHertz QPOs) found in the power spectra
of non-AMSPs are correlated in a similar way between sources
\citep[see, e.g.,][]{Straaten02,Straaten03,Altamirano08}.
It has also been shown that the shape of the power spectra between
AMSPs and non-AMSPs are very similar.
However, the frequency correlations for AMSPs and non-AMSPs are found
to be shifted in frequency \citep{Straaten05,Linares05} and only one
non-AMSP (4U~1820--30) might show frequency shifts as those seen in
AMSPs \citep{Altamirano05}. 
Confirming whether the pulse mechanism in AMSPs is related or not with
these frequency shifts might give important clues that help us
understand the differences between AMSPs and non-AMSPs.

\begin{figure}[!hbtp] 
\resizebox{1\columnwidth}{!}{\rotatebox{0}{\includegraphics[clip]{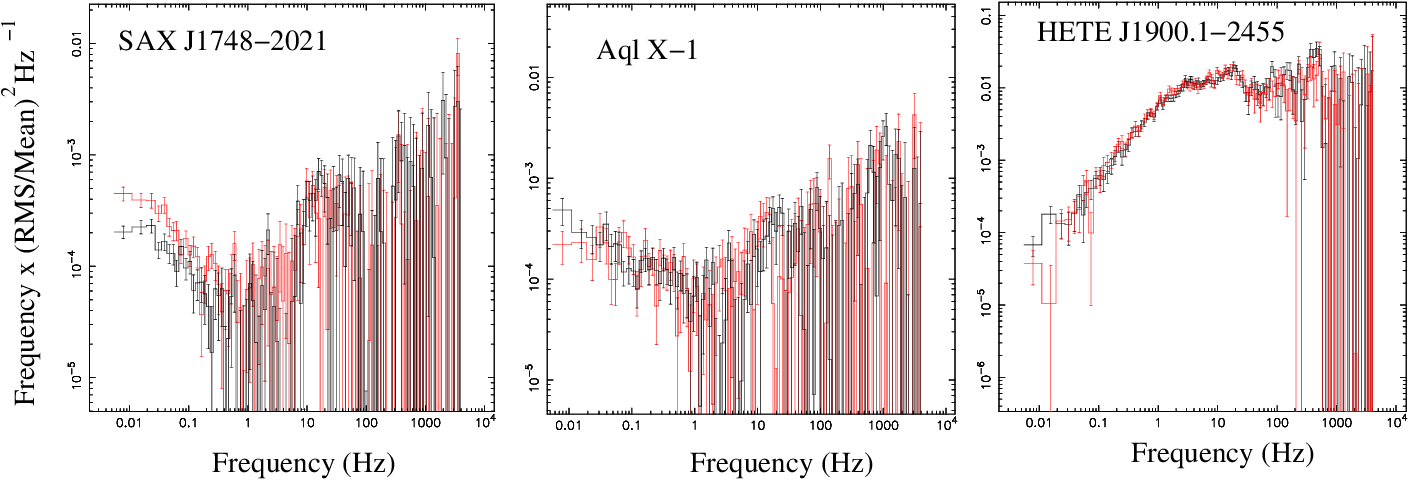}}}
\caption{Representative power spectra during periods with and without
  pulsations (black and gray, respectively) for SAX J1748--2021 (left),
  Aql X-1 (middle) and HETE~J1900.1-2455 (right).  }\label{fig:pds}
\end{figure}

The intermittent AMSPs now allow us to further investigate differences
in the aperiodic variability during pulsating and non-pulsating
intervals. 
As a first step, we analyzed all public observations of these three
sources, and compare power spectra within each source. We find that
(i) for SAX J1748--2021 most of the power spectra are typically of the
(upper) banana state, during both pulsating and non-pulsating periods;
the low-frequency noise seems to be weaker when pulsations are
present;
(ii) the average power spectrum of the observation of Aql X-1 in which
pulsations were detected is typical of the upper banana state; the
statistics is not enough to allow any detailed study of the shape of
the PDS during the 150 seconds in which pulsations were detected; 
(iii) most of the power spectra of HETE~J1900.1-2455 are like those
typically seen during the EIS and IS. We found identical (within
errors) power spectra during the pulsating and the non-pulsating
periods.
In Figure~\ref{fig:pds} we show representative power spectra. Our
results suggest that frequency correlations within a given
intermittent source will not differ between pulsating and
non-pulsating periods.

\section{Questions and Answers}

Only 10 out of more than 100 neutron stars LMXBs have shown
millisecond pulsations associated with the spin frequency. Why these
sources are different has been the subject of many discussions during
the last decade.
Now, the recent discovery of the 3 intermittent AMSPs allow us to
better study differences between sources, as well as to conclude that
irrespective of the mechanisms behind the pulsations in these three
sources, it is now clear that there is not a strict division between
pulsating and non-pulsating sources. It is possible that all sources
pulsate occasionally although the recurrence times could be very long.

Although the intermittent AMSPs might give us new clues to understand
these systems, they also raise new questions:
can a unique scenario explain all the intermittency observed
(HETE~J1900.1-2455 shows a decrease in pulse amplitude in timescales
of days or weeks, while Aql X-1 and SAX~J1748--2021 in timescales of
hundreds of seconds)?
Is there a relation between the pulse mechanism and the thermonuclear
burning on the neutron star surface? 
What is the relation between the pulse mechanism and the burst
oscillation mechanism? Interestingly, bursts and burst oscillations
have been detected in the AMSPs SAX~J1808.4-3658 and XTE~J1814-338,
and bursts without burst oscillations in the two intermittent AMSPs
SAX~J1748-2021 and HETE~J1900.1-2455. The only connecting system might
be Aql X-1, since no X-ray bursts have been detected in the other
AMSPs.
Does the pulse mechanism affect in any way the observed aperiodic
variability?
Some of these questions are discussed in this volume.

\bibliographystyle{aipproc}   % if natbib is available
%bibliography{../Most_complete_bib}

%\bibliography{sample}

%  \typeout{******************************************}
%  \typeout{** Please run "bibtex \jobname" to optain}
%  \typeout{** the bibliography and then re-run LaTeX}%
%  \typeout{** twice to fix the references!}
%  \typeout{******************************************}

\end{document}